\documentclass[12pt]{article}
\usepackage{latexsym}
\usepackage{amssymb}
\usepackage{amsmath}
\usepackage{graphicx}
\usepackage{enumitem}
\usepackage{slashed}
\usepackage{hyperref}

\def\be{\begin{equation}}
\def\ee{\end{equation}}
\def\bear{\begin{eqnarray}}
\def\eear{\end{eqnarray}}
\def\nn{\nonumber}

\newcommand\bra[1]{{\langle {#1}|}}
\newcommand\ket[1]{{|{#1}\rangle}}

\def\s{\sigma}

\def\bra{\langle}
\def\ket{\rangle}

\newcommand{\ti}[1]{\tilde{#1}}

\newcommand{\sm}[1]{\mbox{\scriptsize #1}}
\newcommand{\tn}[1]{\mbox{\tiny #1}}
\renewcommand{\@}[1]{\sqrt{#1}}
\renewcommand{\le}[1]{\label{#1}\end{eqnarray}}
\newcommand{\bea}{\begin{eqnarray}}
\newcommand{\eea}{\end{eqnarray}}

\newcommand{\eq}[1]{(\ref{#1})}
\def\nn{\nonumber\\}

\def\ffract#1#2{\raise .35 em\hbox{$\scriptstyle#1$}\kern-.25em/
\kern-.2em\lower .22 em \hbox{$\scriptstyle#2$}}

\begin{document}

\pagestyle{empty}

\centerline{{\Large \bf On Empirical Equivalence and Duality}}
\vskip.7cm

\begin{center}
{\large Sebastian De Haro}\\
\vskip .7truecm
{\it Trinity College, Cambridge, CB2 1TQ, United Kingdom}\footnote{To appear in {\it Hundred Years of Gauge Theory,} S.~De Bianchi and C.~Kiefer (Eds.), Springer, for the centenary of the publication of Hermann Weyl's {\it Raum-Zeit-Materie}.}\\
{\it Department of History and Philosophy of Science, University of Cambridge}\\
{\it Vossius Center for History of Humanities and Sciences, University of Amsterdam}

\vskip .7truecm
{\tt sd696@cam.ac.uk}
\end{center}

\vskip1cm
\begin{quote}
`A science can never determine its subject-matter except up to an isomorphic representation' (Hermann Weyl, 1934).
\end{quote}
\vskip 1cm
\begin{center}
\today
\end{center}

\vskip 7truecm

\begin{center}
\textbf{\large \bf Abstract}
\end{center}

I argue that, on a judicious reading of two existing criteria---one syntactic and the other semantic---dual theories can be taken to be empirically equivalent. The judicious reading is straightforward, but leads to the surprising conclusion that very different-looking theories can have equivalent empirical content. And thus it shows how a widespread scientific practice, of interpreting duals as empirically equivalent, can be understood by a thus-far unnoticed feature of existing accounts of empirical equivalence.

\newpage
\pagestyle{plain}

\tableofcontents

\newpage

\section{Introduction}\label{intro}

The phenomenon of duality has been a central topic in theoretical physics for several decades. A duality is an isomorphism between (possibly very different-looking) theories, and so dualities are powerful tools for theory construction.\footnote{Isomorphisms, in a sense different to the one used here, are also discussed in the literature on scientific representation. On the semantic view, models represent their target systems in virtue of their being isomorphic to them (other morphisms are sometimes also used). See for example French (2003) and Frigg and Nguyen (2016). For a discussion in the context of ontic structural realism, see French and Ladyman (1999:~p.~108).} They are also very useful in describing empirical phenomena that would otherwise be intractable.

The main question that I will address in this paper is whether dual theories are, or can be taken to be, empirically equivalent---as physicists often claim that they are---on the existing criteria of empirical equivalence. Given that a duality in physics is, in essence, an isomorphism between theories, this question reminds us of Weyl's question of whether physics can determine its own subject-matter---to which he answered that this is only possible up to isomorphism.\footnote{Weyl (1934:~pp.~95-96); see also p.~129.}

My motivation is two-fold. First---as I will argue in a moment---empirical equivalence is an aspect of dualities that has remained relatively under-developed in the recent philosophical discussions, compared to the rich extant analyses of theoretical equivalence (sometimes also called `physical equivalence'); and yet it is often presupposed by those analyses. Second, physicists commonly say that dual theories are empirically equivalent, and standardly use this in their theoretical constructions: so that, without an account that explains how duals can be empirically equivalent, this important scientific practice would be unintelligible.\footnote{Empirical equivalence is also essential in discussions of empirical under-determination in connection with scientific realism. I take this up in De Haro (2020): and so, I will not pursue it further here.}

About the first motivation: while the recent philosophical discussion of dualities has focussed on the analysis of theoretical equivalence, it has for the most part {\it assumed} that dual theories are empirically equivalent, without normally being explicit about a detailed criterion of empirical equivalence, and without arguing that, in general, two duals are empirically equivalent under that criterion.\footnote{Arguments for empirical equivalence have of course been given in specific examples. For electric-magnetic duality, see Dieks et al.~(2015:~pp.~209-210) and Weatherall (2019:~Section 3); for T duality, see Huggett (2017) and Butterfield (2018:~Section 6.3); for gauge-gravity dualities, see De Haro (2017:~pp.~116-117) and  Dawid (2017:~pp.~24-26).} For example, Fraser (2017:~p.~1) writes that `dual theories are regarded as not merely empirically equivalent, but physically equivalent', and discusses an example of a formal equivalence between Euclidean field theory and relativistic quantum field theory, in which `the type of equivalence that obtains... goes well beyond empirical equivalence, but falls short of physical equivalence' (p.~2). Also, Read (2016:~p.~213) and Le Bihan and Read (2018:~p.~2) cite van Fraassen's (1980) notion of empirical equivalence, and take empirical equivalence as a necessary condition for two theories to be dual. And, while I agree with these papers, I also suggest that one needs to explain {\it how} dual theories can be empirically equivalent, according to the standard criteria of empirical equivalence---or to explain how these criteria need to be modified, in order to get empirical equivalence of dual theories. In other words, empirical equivalence is not automatic, and needs to be argued for.\footnote{For a discussion of empirical equivalence in the context of category theory, see Weatherall (2019). Although Weatherall appears to stress the lack of empirical equivalence more than I do, we both argue that ``duals are not automatically empirical equivalent'', and we both conclude that it is nevertheless possible to view dual theories as empirically equivalent. The present paper, specifically, works out how two duals can be empirically equivalent on the syntactic and semantic conceptions. Another author worth mentioning is Dawid (2017:~p.~28), who argues that empirical equivalence takes on a different role in string theory than it traditionally does in the scientific process: namely, as an indicator of important constraints on theory construction, that he argues is not fully visible in the classical limit. For a discussion of Dawid's views in relation to mine, see De Haro (2020:~Section 4.2).}

About the second motivation: it would be desirable to have an account of empirical equivalence that can be applied more generally to modern theories of physics. To this end, there are two broad traditional accounts of the notion of empirical equivalence in the philosophical literature that are useful: of which one is syntactic (see for example Quine 1970, 1975) and the other semantic (see for example van Fraassen 1980, 1989). Thus it would be of considerable interest to see whether or not they give the same verdicts for dualities. For example, some---though not all---recent accounts of theoretical equivalence, notably Weatherall (2016, 2019), give verdicts of empirical equivalence in specific examples, but without making explicit the connection with these accounts of empirical equivalence. %\footnote{While the older literature on theoretical equivalence, for example Quine (1975) and Glymour (***), took empirical equivalence to be a prerequisite for theoretical equivalence, some of the recent literature, such as Barrett and Halvorson (2016), do not mention empirical equivalence or interpretation at all---their criterion of theoretical equivalence is presented as being 'formal', and interpretation is in terms of formulas in a formal language.} 
And thus it seems important, for these accounts of theoretical equivalence and for modern theories of physics generally, to see whether the syntactic and semantic accounts of empirical equivalence involve any significant differences.

Thus my project starts with a comparison of the criteria of empirical equivalence that are given by the syntactic and semantic conceptions of theories. I will do this in Lutz's (2017) spirit of reconciliation between the two: indeed my argument will be that, although the two criteria of empirical equivalence give {\it prima facie} different verdicts in various cases, on a deeper---and perhaps surprising---analysis they can be reconciled. The analysis in question is suggested by dualities.

At first sight, the syntactic criterion of empirical equivalence appears to be stronger than the semantic criterion because, while the former requires {\it identity} of observational sentences, the latter requires only {\it isomorphism} of models. Indeed, this distinction between identity and isomorphism is not innocuous: and it partly {\it motivates} van Fraassen's (1980) adoption of the more liberal semantic, as against the stricter syntactic, criterion. And since dualities are isomorphisms rather than identities, dual theories are empirically {\it in}equivalent by a straightforward application of the syntactic criterion: while empirically equivalent by the semantic criterion. In Section \ref{eed}, I will motivate the {\it judicious reading} of van Fraassen's (1980, 1989) criterion which leads to this verdict, as a surprising but straightforward new application of his proposal.

But, as I announced, the two views can be reconciled: and this entails a new application of the {\it syntactic} criterion of empirical equivalence, to a {\it different pair of theories}: namely, a theory $T$ and its ``reinterpreted dual'', $T'$, such that these two theories---one of which is reinterpreted using the duality---are empirically equivalent. 

Thus the semantic and syntactic criteria can be made to give the same verdicts: through a judicious reading of the semantic criterion, and applying the syntactic criterion to a different pair of theories. 

Underlying this possibility of using the duality to ``change the interpretation'' is the idea that formal theories admit various interpretations, and that the discovery of a duality can be a good reason to reinterpret a given theory---especially if  such a reinterpretation is motivated by scientific practice. This, then, not only agrees with, but also {\it explains,} a fruitful and widespread scientific practice. Indeed, I believe that a conception of the criteria of empirical equivalence that did not allow for dual theories to be empirically equivalent would render the importance of dualities, in current theoretical physics, philosophically unintelligible.

In Section \ref{emeq}, I review Quine's syntactic and van Fraassen's semantic conceptions of empirical equivalence. Then, in Section \ref{secdual}, I briefly introduce duality, and illustrate it in some examples. Section \ref{eed} brings the two topics together, and finds that dual theories can---surprisingly---be taken to be empirically equivalent. Section \ref{conclusion} summarises the paper's main thesis.

\section{Empirical Equivalence}\label{emeq}

In this Section, I review the two criteria of empirical equivalence that I will compare in Section \ref{eed}: Quine's (1970, 1975) syntactic, and van Fraassen's (1980, 1989) semantic criterion.

Quine (1975:~p.~319) says that two theories are {\it empirically equivalent} if they imply the same observational sentences---also called `observational conditionals'---for all possible observations: present, past, future\footnote{Quine (1970:~p.~179).} or `pegged to inaccessible place-times'.\footnote{Quine (1975:~p.~234). One area where inaccessible place-times appear is of course cosmology. See Glymour (1977), Malament (1977) and Manchak (2009), who discuss the under-determination of topology by local geometric structure.\label{udc}} He puts it thus:
\begin{quote}\small The empirical content of a theory formulation is summed up in the observation conditionals that the formulation implies (Quine 1975:~p.~323).
\end{quote}
Quine does not tell us, in either of his two papers on empirical under-determination (1970, 1975), what he means by `observation'. But there are of course a few general things one can say. First, his views on observation were formed against the background logical positivist view that observation ultimately reduces to human sense data. However, his views would not have entailed stronger logical positivist doctrines such as Carnap's reduction of theoretical terms to observational terms. He summarises  his empiricist commitments thus: `Two cardinal tenets of empiricism remained unassailable... and so remain to this day. One is that whatever evidence there {\it is} for science {\it is} sensory evidence. The other... is that all inculcation of meanings of words must rest ultimately on sensory evidence' (2002:~p.~249). His mention of `inaccessible place-times' (1975:~p.~234) suggests that, by observation, he had something broader in mind than mere perception by the human senses---but what, he does not say. 

Another influential account of the meaning of `empirical' is by van Fraassen (1980:~p.~64): 

\begin{quote}\small To present a theory is... to present certain parts of those models (the {\it empirical substructures}) as candidates for the direct representation of observable phenomena. The structures which can be described in experimental and measurement reports we call {\it appearances}: the theory is empirically adequate if it has some model such that all appearances are isomorphic to empirical substructures of that model. 
\end{quote}
Van Fraassen famously restricts the scope of `observable phenomena' to observation by the unaided human senses. Accordingly, his mention of `experimental and measurement reports' is restricted in the kinds of experiments and measurements that it affords. Thus I will set van Fraassen's notion of observability aside but keep his notion of empirical adequacy as a useful {\it semantic} alternative to Quine's {\it syntactic} construal of the empirical.\footnote{For an account of observation that, in my opinion, resonates better with modern science, see Lenzen (1955).}

Van Fraassen's (1980:~p.~67) notion of {\it empirical equivalence} also involves consideration of the theory's {\it models:} namely, the structures that satisfy the theorems of the theory; alternatively, the structures that comprise the theory, regarded as a collection of models. It is the {\it empirical substructures} of those models that `are candidates for the direct representation of observable phenomena' (1980:~p.~64). Thus two theories, $T$ and $T'$, are {\it empirically equivalent} if for every model $M$ of $T$ there is a model $M'$ of $T'$ such that all empirical substructures of $M$ are isomorphic to empirical substructures of $M'$, and vice-versa.

To summarise: two theories are {\bf empirically equivalent} if they imply {\it the same observational sentences} (on Quine's syntactic conception) or if {\it the empirical substructures of their models are isomorphic to each other} (on van Fraassen's semantic conception). %In both cases, I take `observable' to mean `accessible through physical interaction'.

\section{Duality}\label{secdual}

This Section introduces the notion of duality, and gives two examples that I will use in my analysis of empirical equivalence in Section \ref{eed}.\footnote{A detailed Schema for dualities is presented in De Haro (2016) and De Haro and Butterfield (2018). Further work on dualities is in Dieks et al.~(2015), Huggett (2017), Read and M\o ller-Nielsen (2018), and Rickles (2017).} 

A duality is an isomorphism between two theory formulations. I will denote the duality map by $d:T\rightarrow T'$, where $T$ and $T'$ are two bare theories, i.e.~before we give them a physical interpretation. The duality maps (isomorphically) the states of $T$ to the states of $T'$, and likewise for their quantities: while preserving the {\it values} of the quantities, the dynamics, and the symmetries that are stipulated for $T$ and $T'$.\footnote{For the relation between symmetry and duality, see De Haro and Butterfield (2019).}

An interpretation can be modelled using the idea of an {\it interpretation map}: namely, a partial function mapping bits of a bare theory (paradigmatically: the states and the quantities) into the theory's domain of phenomena. %, i.e.~appropriate properties and relations in the physical world. 
I will denote such a map by: $i:T\rightarrow D$.\footnote{For a detailed exposition of interpretations in terms of maps, the conditions they satisfy, and how this can be used for referential semantics semantics, see De Haro and Butterfield (2018) and  De Haro (2016, 2017). }

Consider, for example, orthodox quantum mechanics, as often presented in textbooks. Its interpretation map, $i$, maps the bare theory to its domain of phenomena as follows:
\bea
i(x)&=&\mbox{\small`the position, with value $x$, of the particle upon measurement'}\nn
i\left(|\psi(x)|^2\right)&=&\mbox{\small`the probability density of finding the particle at position $x$, upon measurement'}\nonumber
%\nn
%i(\psi)&=&\mbox{\small`the physical state of the system (the particle)'},\nonumber
\eea
etc., where $x$ is an eigenvalue of the position operator. Thus %wave-functions $\psi$ (the states of the bare theory) are interpreted as physical states of systems. S
the eigenvalues of self-adjoint operators are interpreted as %values of properties of a physical system, such as `energy', `spin', etc. 
%Eigenvalues of self-adjoint operators are interpreted as 
possible outcomes of individual measurements. % that determine the values of determinate properties of the system. 
The absolute value squared of wave-functions are interpreted as Born probabilities, etc. 

Notice that not everything that is contained in the domain $D$ (such as: outcomes of measurements, Born probabilities, physical states of a system, etc.) counts as `empirical'. The values of quantities for a given state are typically observable, but the states themselves are not. This motivates the identification, for quantum mechanics, of van Fraassen's {\it empirical substructures} with the set of transition amplitudes of self-adjoint operators, and the expressions in which they appear.  \\

\begin{figure}
\begin{center}
\includegraphics[height=3.5cm]{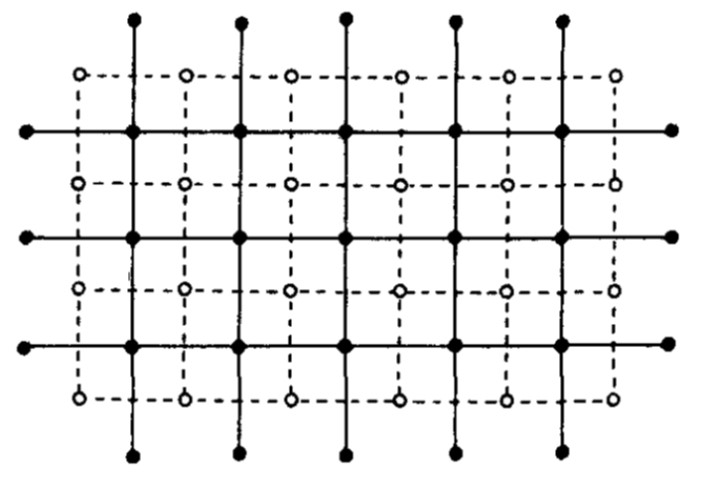}~~~~~~~~~~~~~~
\includegraphics[height=3.5cm]{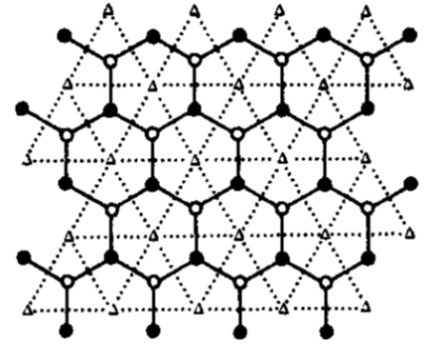}
\caption{\small A lattice (solid lines) and its dual lattice (broken lines). Left: square lattice. Right: honeycomb lattice.
Baxter (1982), with permission.}
\label{Ising}
\end{center}
\end{figure}

I will llustrate {\it duality} in the example of Kramers-Wannier duality in statistical mechanics. Consider the Ising model on a square lattice (see Figure \ref{Ising}).\footnote{I follow the treatment in Baxter (1982:~Section 6.2) and Savit (1980:~pp.~456-457).} 
Each lattice site, i.e.~each vertex of the lattice, is occupied by a spin $\s_i$ (where $i$ labels the lattice sites) with two possible values: $+1$ or $-1$. Two nearest-neighbour spins $\s_i$ and $\s_j$ contribute a potential energy $-J\,\s_i\,\s_j$, where $J$ is some fixed energy, and the total energy is the sum over all such pairs. 

The {\it partition function} is defined as the sum of the exponentials of the energy (Boltzmann factors), summed over all the states. Thus for a square lattice of $N$ sites, we sum the exponentials of the energies of the pairs:
\bea
Z_N:=\sum_\s e^{-E(\s)/k_{\tn B}T}=\sum_\s\exp\left(K\sum_{(i,j)}\s_i\,\s_j\right)=:e^{-N\,F_N}~,\nonumber
\eea
where we sum over pairs $(i,j)$, i.e.~all {\it edges} (nearest-neighbour pairs of spins). The sum over $\s$ is the sum over all spin values, $\pm1$. Furthermore, I have defined $K:=J/k_{\tn B}T$, where $k_{\tn B}$ is Boltzmann's constant and $T$ is the temperature. On the right is the definition of the free energy, $F_N$.

Kramers-Wannier duality involves a reinterpretation of the spin variables, as belonging to the `dual lattice', which one gets by placing points at the centres of the faces of the original lattice, and connecting the points in adjacent faces (i.e.~faces that share an edge): see Figure \ref{Ising}. The result, for the square lattice, is that the original Ising model (with weight $K$) and the Ising model on the dual lattice (with weight $K^*$) are related to each other in the limit that the lattice is infinite, i.e.~$N\rightarrow\infty$. First, one defines the free energy in the limit of infinite $N$:
\bea
{\cal F}(K):=\lim_{N\rightarrow\infty}\,F_N=-\lim_{N\rightarrow\infty}~{1\over N}\,\ln Z_N~,
\eea
and then one derives the following tranformation property of the free energy:
\bea\label{Isingd}
{\cal F}(K^*)&=&{\cal F}(K)+\ln\sinh2K\\
\tanh K^*&:=&e^{-2K}~.\nonumber
\eea
Thus the free energies of the two models are equal, up to $\ln\sinh2K$. Notice that the $K\rightarrow \infty$ limit of one model maps, through Eq.~\eq{Isingd}, to the $K^*\rightarrow0$ limit of the other, and vice-versa. Alternatively, since $K$ and $K^*$ depend on the temperatures $T$ and $T^*$, respectively, {\it the Ising model at high temperature is dual to the Ising model at low temperature, and vice-versa.}

This is the basic idea of Kramers-Wannier duality: there is a well-defined one-to-one map between the two models (the high-temperature and the low-temperature Ising models, with their specific weights) such that the free energies of the two models and all other quantities map onto each other, through Eq.~\eq{Isingd}. Furthermore, this map generalises to other quantities of interest, such as the correlation functions, $\bra\s_i\,\s_l\ket$, between spins $\s_i$ and $\s_l$ that may be arbitrarily far away from each other (see Savit 1980:~p.~457). If this map generalises to all the quantities of interest in the theory, which depend on arbitrary spin states, then it is an isomorphism between the two theories---it is a {\it duality.}\\

My second example of a duality is gauge-gravity duality, which was used in the RHIC experiments in Brookhaven, NY. The duality successfully relates the four-dimensional quantum field theory (QCD, quantum chromodynamics) that describes the quark-gluon plasma, produced in high-energy collisions between lead atoms, to the properties of a five-dimensional black hole. The latter was employed to perform a calculation that, via an approximate duality, provided a result in QCD: namely, the shear-viscosity-to-entropy-density ratio of the plasma, which could not be obtained in the theory of QCD  describing the plasma. Thus a five-dimensional black hole is used to describe, at least approximately, an entirely different (four-dimensional!) empirical situation.

\section{Empirical Equivalence of Dual Theories}\label{eed}

In this Section, I will apply the syntactic and semantic criteria of empirical equivalence, from Section \ref{emeq}, to cases of duality. The application will involve a judicious reading of these criteria.

My `judicious reading' is liberal but also straightforward, and it has two motivations. First, it captures an important scientific practice of using dualities to construct new theories that describe empirically equivalent situations, as in the RHIC experiment, discussed at the end of the previous Section. 
Second, the judicious reading is independently motivated by a historical-critical analysis of van Fraassen's semantic criterion of empirical equivalence. And I will claim that it casts light on some of the alleged differences between the semantic and syntactic views of theories.

On the syntactic criterion, two theories are empirically equivalent if they imply the same observational sentences. Since a duality is an isomorphism, $d:T\rightarrow T'$, between two theories whose domains of phenomena can be very different, dual theories imply different observational sentences, and are in general {\it not} empirically equivalent in this sense. On their ordinary interpretations,\footnote{Dieks et al.~(2015) and De Haro (2016, 2017) call this an `external' interpretation, where the meaning of the terms is fixed from outside the theory. The reinterpretations discussed below are `internal' interpretations, which take the duality as a starting point for establishing the meanings of the terms.} 
QCD and the five-dimensional gravitational theory make different predictions, even though the numerical values of their quantities agree. And under Kramers-Wannier duality, a {\it high-temperature} lattice maps to a {\it low-temperature} lattice (alternatively, strong coupling $K$ is mapped to weak coupling $K^*$, according to Eq.~\eq{Isingd}). 
Thus their observational sentences differ.

On the semantic criterion, two theories are empirically equivalent {\it if the empirical substructures of their models are isomorphic to each other} (cf.~the end of Section \ref{emeq}). 

Notice that it is `isomorphism' of the empirical substructures of the theory's models, rather than `identity', that counts here---and I will argue in Section \ref{vFee} that this literal reading of van Fraassen is indeed correct. 
Thus let me take van Fraassen's quote from Section \ref{emeq} literally, and look for a suitable {\it isomorphism} between the empirical substructures of the models that we consider. Since we are dealing with dualities, the suggestion is that the duality map gives a natural---though surprising---new candidate for such an isomorphism: which I will dub the `induced duality map'.

Thus consider the case in which the dual theories' domains of phenomena are distinct but isomorphic, according to an `induced duality map', $\ti d:D_1\rightarrow D_2$. The commuting diagram in Figure \ref{dtilde} will not always close (the condition for its closure is that $i_2\,\circ\,d=\ti d\,\circ\,i_1$; cf.~De Haro, 2019:~Section 2.2.3). But if it does, then the two theories are clearly {\it empirically equivalent} on van Fraassen's conception taken literally.
\begin{figure}
\begin{center}
\bea
\begin{array}{ccc}T_1&\overset{d}{\longleftrightarrow}&T_2\\
~~\Big\downarrow {\sm{$i_1$}}&&~~\Big\downarrow {\sm{$i_2$}}\\
D_1&\overset{\ti d}{\longleftrightarrow}&D_2
\end{array}\nonumber
\eea
\caption{\small Empirical equivalence. There is an induced duality map, $\ti d$, between the domains.}
\label{dtilde}
\end{center}
\end{figure}

For Kramers-Wannier duality, the map $\ti d$ replaces a lattice by its dual, and translates the value of the temperature from one to the other. In the case of the RHIC experiments, the calculations are done in the five-dimensional theory, and then they are translated (using the induced duality map) into predictions about the four-dimensional plasma.

Thus on this literal reading of van Fraassen, the dualities that we have discussed {\it do} in fact (and surprisingly!) relate empirically equivalent theories, by a reinterpretation of their domains of phenomena. What {\it prima facie} looks like a five-dimensional black hole can, through the induced duality map, be reinterpreted as a four-dimensional plasma.

The same holds of course if we consider Newtonian mechanics with different standards of rest (an example that van Fraassen himself considers). According to Newton and Clarke, different standards of rest give different empirical situations: but they are {\it isomorphic} situations, because any standard of rest can be mapped to any other by a Galilei transformation. And so, different standards of rest give empirically equivalent situations on van Fraassen's criterion (as in Leibniz).\footnote{For a discussion of this example as a case of duality, see Butterfield (2019) and De Haro and Butterfield (2019).}

Thus van Fraassen's semantic criterion---because it involves isomorphism, rather than identity, of substructures---seems more liberal than the syntactic criterion. By Quine's criterion, the examples of dualities are not cases of under-determination; by van Fraassen's, they are. This prompts two questions, the first of which leads to the second:\\

(A)~~Is my interpretation of van Fraassen's notion of empirical equivalence correct? Does his notion allow for isomorphisms that involve dualities?

(B)~~Can the syntactic notion of empirical equivalence also allow the same latitude? 

\subsection{Van Fraassen on empirical equivalence}\label{vFee}

As to question (A): the evidence that my interpretation is straightforward is from van Fraassen's (1980, 1989). First, in his famous example of the seven point geometry (1980:~p.~43) that motivates the semantic notion of a `model', he writes that `the seven-point structure can be {\it embedded} in a Euclidean space. We say that one structure can be embedded in another, if the first is isomorphic to a part (substructure) of the second'. In (1989:~pp.~219-220), he adds:  
`This relation [of isomorphism] is important because it is also the exact relation a phenomenon bears to some model of a theory, if that theory is empirically adequate.' Why does van Fraassen use---even emphasise---`embedding' and `isomorphism',\footnote{Van Fraassen (1980, 1989) is constant in his use of `isomorphism' in connection with empirical equivalence.} rather than just saying that `the seven-point structure... is {\it equal to,} or is {\it the same as,} a part of Euclidean space'? Like isomorphism, an embedding is a mathematical notion: an embedding essentially comes down to an injective map, such that when restricted to the image set of the map, the map thus obtained is an isomorphism.\footnote{For a typical definition in the context of topology, see for example Munkres (2000:~p.105).} His use of both `isomorphism' and `embedding', rather than the more straightforward `equality', or `sameness', implies that he {\it means} isomorphism and embedding: so that the literal reading is correct---and his second quote clarifies that isomorphism is important not only because this is how models relate to one another, but also with their empirical adequacy. 

Second, van Fraassen (1980:~pp.~45-50) discusses the empirical adequacy of Newton's theory of mechanics and gravitation, and the {\it empirical equivalence} of the alternatives to this theory with the additional postulate that the centre of gravity of the solar system has constant absolute velocity $v$ (he denotes these theories by `$\mbox{TN}(v)$'): 

\begin{quote}
\small When Newton claims empirical adequacy for his theory, he is claiming that his theory has some model such that {\it all actual appearances are identifiable with (isomorphic to) motions} in that model' (p.~45, his italics). 
\end{quote}
The use of `isomorphism' is here essential: 
\begin{quote}
\small
all the theories $\mbox{TN}(v)$ are empirically equivalent exactly {\it if all the motions in a model of $\mbox{TN}(v)$ are isomorphic to motions in a model of $\mbox{TN}(v+w)$} (p.~46, his italics).
\end{quote}
%, for all constant velocities $v$ and $w$}' 
There can be no doubt that it is the {\it isomorphism} of the motions between $\mbox{TN}(v)$ and $\mbox{TN}(v+w)$ that makes the theories {\it empirically equivalent,} regardless of the value of $w$. And in so far as one member of the family is empirically  adequate (cf.~p.~47), the whole family is empirically adequate.
(He also argues that Maxwell's theories with different absolute velocities are empirically equivalent.)

But if all {\it these} theories count as empirically equivalent, then we should count other theories that are related by a similar isomorphism, of the state-spaces and quantities (i.e.~dualities), as empirically equivalent. 
%---as I said before, the equivalence between models of Newtonian mechanics with different standards of rest is one example of a duality. 

But there is more evidence in support of my reading. For the semantic view's use of `isomorphism' in its conception of empirical adequacy underpins van Fraassen's (1980:~pp.~53-56) preference for the semantic over the syntactic approach. Van Fraassen claims that the syntactic approach judges the theories $\mbox{TN}(v)$ with different values of the velocity $v$ to be {\it empirically inequivalent,} i.e.~the syntactic approach distinguishes between the empirical consequences of theories for which there should be no such distinction: 

\begin{quote}
\small [On the syntactic approach,] $\mbox{TN}(0)$ is no longer empirically equivalent to the other theories $\mbox{TN}(v)$' (p.~55). 
\end{quote}

Even though van Fraassen here targets an untenable version of the syntactic view (namely, the old logical positivist view that theoretical sentences are reduced to observational sentences), and the syntactic view may well have its own resources to distinguish between versions of Newtonian theory: I agree that the syntactic notion of empirical equivalence is {\it prima facie} stronger, because of the difference between identity and isomorphism---and coincides with my own interpretation of Quine's notion, at the beginning of this Section.\footnote{Van Fraassen (2014), in his reply to Halvorson (2012), has recently emphasised the importance of interpretation for questions of equivalence: he always took interpretative notions to be properly accounted for in the semantic conceptions of theories. As such, this does not contradict any of the above, which explicitly takes interpretation into account. Notice, furthermore, that van Fraassen nowhere says that `empirically equivalent interpretations should have the same truth values' or something of the sort. Indeed, he explicitly says that `if we believe of a family of theories that are all empirically adequate, but each goes beyond the phenomena, then we are still free to believe that each is false' (1980:~p.~47). Thus the empirical adequacy (and hence equivalence) of theories is independent of their truth. Furthermore, van Fraassen (2014) is concerned with Halvorson's discussion of theoretical equivalence more than with empirical equivalence.}
Thus van Fraassen seems to be making the same distinction that I made earlier in this Section. (I disagree that this is a reason to regard either criterion as superior, as I will explain below.)

Finally, let me add a word about why duality (and induced duality) are the right kinds of isomorphism here, i.e.~why I am allowed to generalise van Fraassen's relatively simple isomorphisms to dualities. To see this, one needs to specify what is the relevant `isomorphism': for isomorphism is a notion that is well-defined only relative to a given structure. For scientific theories, the natural isomorphisms to consider are those that relate their structures. And since van Fraassen (1980, 1989) presents scientific theories in terms of states and quantities, with their corresponding interpretation maps (see also 1970:~pp.~329, 334-335), the relevant isomorphisms between scientific theories should relate these structures. But this is precisely what the duality and induced duality maps do: they map the states and the quantities, the values of the quantities. In other words, once one has agreed that a theory is formulated as a structured set of states, quantities, and dynamics, a principled definition of an isomorphism of theories (and models) should preserve {\it those} structures---and duality is such a principled isomorphism. In this light, it is no coincidence that van Fraassen's and the Schema's verdicts about the empirical equivalence of Newtonian theory coincide. 
Thus the claim that dual theories can be taken to be empirically equivalent is a natural (although perhaps unexpected!) application of van Fraassen's semantic criterion.

I have discussed van Fraassen (1980) in some detail because of its wide influence, and to show how his notion of empirical equivalence applies to dualities. I submit that the textual evidence leaves no doubt that my interpretation of dual theories as being empirically equivalent, through an induced duality map $\ti d$, 
is a straightforward application not just of van Fraassen's general notion of empirical equivalence, but indeed of his main motivation for developing a semantic account: namely, that he thinks that such an account {\it must} make isomorphic theories empirically equivalent---as his discussion of Newton's and Maxwell's theories, and his criticisms of the syntactic conception, show.

\subsection{Applying the syntactic notion of empirical equivalence}\label{synee}

As to question (B) above: I claim that the syntactic criterion of empirical equivalence can similarly be applied in a more liberal way, like the semantic criterion. 
And it will not even be necessary to {\it modify} Quine's criterion of empirical equivalence from Section \ref{emeq}; all we need is to apply it to a {\it new pair of theories,} where a new theory is generated through a {\it non-standard interpretation} of the bare theory. 

Let me motivate why we would want to do this, using the example of the quark-gluon plasma, discussed at the end of Section \ref{secdual}. There, the five-dimensional black hole was used to answer empirical questions about a plasma that could not be answered using QCD. The five-dimensional black hole, plus a suitable ``translation'', made this possible. And this scientific practice is justified by my analysis of empirical equivalence. If dual theories could not be taken to be empirically equivalent, dualities would be of little interest for the practicing physicist. It is precisely the fact that one of the dual theories can be used in a different context to make a prediction that is otherwise unattainable, that makes dualities scientifically valuable. 

This kind of isomorphism can be introduced on the syntactic conception just as it is on the semantic conception, by a suitable translation between the sets of observational conditionals, $O_1$ and $O_2$, of the two theories (see Figure \ref{dsynt}). If one finds a translation map $t$ that, added to the syntactic interpretation map $j_1:T_1\rightarrow O_2$, makes true the very same sentences as $j_2:T_2\rightarrow O_2$, so that $t\,\circ\,j_1=j_2\,\circ\,d$, then the two theories are rendered empirically equivalent. In other words, $T_1$, with the {\it non-standard interpretation} $t\,\circ\,j_1$, is empirically equivalent to $T_2$. Thus we do not need to change Quine's criterion: we rather change the theory, by giving it a new (and innovative!) interpretation.\footnote{My proposed use of Quine's notion of empirical equivalence, allowing ourselves to generate a theory with a new interpretation from the consideration of the induced duality map, is in the same spirit in which Lutz claims that the syntactic and semantic conceptions do not differ so much after all. For one of the main problems of the syntactic view is that it seemed highly language-dependent (and van Fraassen aimed to solve this problem by moving to a language-independent semantic view). Now Lutz (2017:~pp.~324-326) follows Glymour and others in looking for more liberal criteria, such as definitional equivalence, that can make different theories `mutually interpretable'. Thus my proposal is that one may consider this to hold not just for the notion of theoretical equivalence, but also for empirical equivalence.\label{Lurzsyn}}

\begin{figure}
\begin{center}
\bea
\begin{array}{ccc}T_1&\overset{d}{\longleftrightarrow}&T_2\\
~~\Big\downarrow {\sm{$j_1$}}&&~~\Big\downarrow {\sm{$j_2$}}\\
O_1&\overset{t}{\longleftrightarrow}&O_2
\end{array}\nonumber
\eea
\caption{\small Translation, $t$, between observational conditionals, $O_1$ and $O_2$, on the syntactic view.}
\label{dsynt}
\end{center}
\end{figure}

But also, these reinterpretations are {\it not forbidden}: for notice that nobody said that we had to stick to a single interpretation in order for a theory to make empirical predictions. Although bare theories may have intended interpretations, assigned to them by history and convenience, nothing in the Quinean notion of empirical equivalence prevents us from generating new theories by reinterpreting the old ones, thus extending the predictive power of a bare theory. Indeed, the exercise is not motivated by philosophical speculation, but by scientific practice. The use of this flexibility in explaining heavy-ion collisions illustrates the scientific importance of the procedure.

Notice that, as on the semantic view, the interpretation thus obtained is non-trivial, and it need not always exist: we are using the theory $T_1$ to produce the observational sentences $O_2$ of theory $T_2$, by giving a non-standard interpretation to $T_1$.

Let me briefly discuss the obvious objection: {\it how can a hot lattice be empirically equivalent to a cold lattice?} A thermometer surely ought to tell the difference? 

But this objection can only be made if we are allowed to do measurements on the system ``from the outside'', i.e.~measurements that are not modelled by the theory. If the measurements are modelled by the theory, then empirical equivalence by definition translates the measurements, which are ordinary physical interactions (see the end of Section \ref{emeq}). Namely, the answer to the objection is the one given to the sceptic of length contraction in special relativity, who asks: {\it if two observers in relative motion make different predictions about the length of a body, surely a measuring stick can say who of them is right?} But of course the measuring sticks are {\it  themselves} contracted under a Lorentz transformation, and so seemingly irreconciliable claims turn out to be empirically equivalent after all. For a discussion of this argument for dualities, see Dieks et al.~(2015:~pp.~209-210). 

Agreed: it is of course {\it not mandatory} to take dual theories to be empirically equivalent. This is clearest in the syntactic conception, where a theory and its dual can always be interpreted according to their ordinary interpretations, rather than non-standard ones. And the duals then simply {\it disagree} about empirical matters. But also on the semantic view this is possible, by adopting non-isomorphic interpretations. This parallels the distinction between external vs.~internal interpretations in Dieks et al.~(2015) and De Haro (2016, 2017): namely, interpretations that ``start from the duality'' vs.~interpretations that are ``independent of the duality''. For more on when it is more appropriate to adopt one sort of interpretation or the other, see De Haro (2016).

Let me briefly compare my discussion of empirical equivalence to other authors who write about dualities. While the recent philosophical discussion of dualities has focussed on the analysis of theoretical equivalence,\footnote{For example, Matsubara (2013:~p.~487) and Read (2016:~p.~213) take empirical equivalence as a necessary condition for two theories to be dual.} it has for the most part {\it assumed} that dual theories are empirically equivalent without an explicit analysis of the conception of empirical equivalence used.\footnote{Read and M\o ller-Nielsen (2018:~\S3.1) cite van Fraassen's notion of empirical equivalence.}
My analysis has aimed to make explicit that the verdict of empirical equivalence of dual theories, on the standard conceptions of empirical equivalence, is not automatic: for it required, in the syntactic view, adopting a non-standard interpretation of a bare theory. And in the semantic view, it involved an unexpected application of the isomorphism criterion, which is widened from the familiar cases to theories whose structures look very different. Furthermore, we have seen that such empirical equivalence is not mandatory. 

\section{Conclusion}\label{conclusion}

Both the semantic and the syntactic views allow for special applications that render dual theories empirically equivalent: although the way in which they do this is slightly different. 

In van Fraassen's semantic conception, duals are rendered empirically equivalent because his notion of empirical equivalence has `isomorphism' built into it---and duality is a natural notion of isomorphism between scientific theories. In this way, two dual bare theories are rendered empirically equivalent under their standard interpretations. 
%That such dual theories should be empirically equivalent is not a Pickwickian interpretation of van Fraassen's proposals, but part of his motivation for developing his version of the semantic view. 
This notion of empirical equivalence is of course faithless to meanings, as van Fraassen admits---but this allows him to conclude that different versions of Newtonian mechanics are empirically equivalent, even though they cannot all be true. And this of course articulates an important practice in science.

On the syntactic view, two dual bare theories are usually empirically {\it in}equivalent under their standard interpretations, but are rendered empirically equivalent if one of the theories is given a {\it non-standard interpretation,} i.e.~using a translation map induced from the duality. This does not require changing Quine's criterion of empirical equivalence: it only requires endowing the same bare theory with a new interpretation. In this way, the same bare theory can be put to different uses, including in other domains of phenomena than it was originally developed for. 

And {\it this} move is motivated by the use of dualities in the predictions for the RHIC experiments on quark-gluon plasma: and by many other such uses of dualities by physicists in solving problems in current theoretical physics. % but also by the further development of Quine's ideas within the syntactic conception, as reviewed by Lutz (2018:~pp.~320-321, 323-325) (see footnote \ref{Lurzsyn}).

\section*{Acknowledgements}
\addcontentsline{toc}{section}{Acknowledgements}

I thank Silvia De Bianchi and Claus Kiefer for their invitation to contribute to this volume, and for their comments on the paper. I also thank Jeremy Butterfield, Nick Huggett, and James Weatherall for conversations about the contents of this paper. I also thank John Norton for a discussion of duality and under-determination. This work was supported by the Tarner scholarship in Philosophy of Science and History of Ideas, held at Trinity College, Cambridge.

\section*{References}
\addcontentsline{toc}{section}{References}

\small

Baxter, R.~J.~(1982). {\it Solved Models in Statistical Mechanics}.  Academic Press.\\
\\
Butterfield, J.~(2018). `On Dualities and Equivalences Between Physical Theories'. Forthcoming in {\it Philosophy Beyond Spacetime}, W\"uthrich, C., Huggett, N.~and Le Bihan, B.~(Eds.). Oxford: Oxford University Press.\\
\\
Dawid, R.~(2017). `String Dualities and Empirical Equivalence'. {\it Studies in History and Philosophy of Modern Physics,} 59, pp.~21-29.\\
\\
De Haro, S.~(2016). `Spacetime and Physical Equivalence'. Forthcoming in {\it Beyond Spacetime. The Foundations of Quantum Gravity}, Huggett, N., W\"uthrich, C.~and Matsubara, K.~(Eds.). Cambridge: Cambridge University Press.\\% http://philsci-archive.pitt.edu/13243.\\
%%CITATION = ARXIV:1707.06581;%%
\\
De Haro, S.~(2017). `Dualities and Emergent Gravity: Gauge/Gravity Duality'. {\it Studies in History and Philosophy of Modern Physics,} 59, pp.~109-125.\\
\\
De Haro, S.~(2019). `Theoretical Equivalence and Duality'. {\it Synthese,} topical collection on Symmetries. M. Frisch, R. Dardashti, G. Valente (Eds.), 2019, pp. 1-39. \\
% %%CITATION = ARXIV:1906.11144;%%
\\
De Haro, S.~(2020). `The Empirical Under-determination Argument Against Scientific Realism for Dual Theories'. Forthcoming in {\it Erkenntnis}.\\
\\
De Haro, S.~and Butterfield, J.N.~(2018). `A Schema for Duality, Illustrated by Bosonization'. In: Kouneiher, J.~(Ed.), {\it Foundations of Mathematics and Physics one century after Hilbert}. Springer. \\ %http://philsci-archive.pitt.edu/13229.\\
%%CITATION = ARXIV:1707.06681;%%
\\
De Haro, S.~and Butterfield, J.~N.~(2019). `On Symmetry and Duality'. {\it Synthese,} topical collection on `Symmetries and Asymmetries in Physics', pp. 1-41. Editors: M.~Frisch, R.~Dardashti, G.~Valente. \\
\\
Dieks, D., Dongen, J. van, Haro, S. de~(2015). `Emergence in Holographic Scenarios for Gravity'. {\it Studies in History and Philosophy of Modern Physics} 52 (B), 2015, pp.~203-216.\\ %doi:~10.1016/j.shpsb.2015.07.007.\\
  %%CITATION = ARXIV:1501.04278;%% 
\\
French, S.~(2003). `A Model-Theoretic Account of Representation (Or, I Don't Know Much About Art... But I Know It Involves Isomorphism)'. {\it Philosophy of Science,} 70 (5), pp.~1472-1483.\\
\\
French, S.~and Ladyman, J.~(1999). `Reinflating the Semantic Approach'. {\it International Studies in Philosophy of Science,} 13 (2), pp.~103-121.\\
\\
Frigg, R.~and Nguyen, J.~(2016). `Scientific Representation', {\it Stanford Encyclopedia of Philosophy,} https://plato.stanford.edu/entries/scientific-representation.\\
\\
Glymour, C.~(1977). `The epistemology of geometry'. {\it No$\hat u$s}, pp.~227-251.\\
\\
Halvorson, H.~(2012). `What Scientific Theories Could Not Be'. {\it Philosophy of Science}, 79, pp.~183-206.\\
\\
Huggett, N.~(2017). `Target space $\neq$ space'. {\em Studies in History and Philosophy of Modern Physics}, 59, 81-88. doi:10.1016/j.shpsb.2015.08.007.\\
\\
Fraser, D.~(2017). `Formal and physical equivalence in two cases in contemporary quantum physics'. {\it Studies in History and Philosophy of Modern Physics}, 59, pp.~30-43. \\%doi:~10.1016/j.shpsb.2015.07.005.\\
\\
Le Bihan, B.~and Read, J.~(2018). `Duality and Ontology'. {\it Philosophy Compass,} 13:e12555, pp.~1-15.\\
\\
Lenzen, V.~F.~(1955). `Procedures of Empirical Science'. In: {\it International Encyclopedia of Unified Science,} Neurath, O., Bohr, N., Dewey, J., Russell, B., Carnap, R., and Morris, C.~W.~(Eds.). Volume I, pp.~280-339.\\
\\
Lutz, S.~(2017). `What Was the Syntax-Semantics Debate in the Philosophy of Science About?' {\it Philosophy and Phenomenological Research}, XCV (2), pp.~319-352.\\
\\
Malament, D.~(1977). `Observationally Indistinguishable Space-Times: Comments on Glymour's Paper'. In: {\it Foundations of Space-Time Theories,} J.~Earman, C.~Glymour, J.~Stachel (Eds.). Minnesota Studies in the Philosophy of Science, volume VIII. Minneapolis: University of Minnesota Press.\\
\\
Manchak, J.~B.~(2009). `Can we know the global structure of spacetime?' {\it Studies in History and Philosophy of Modern Physics,} 40, pp.~53-56.\\
\\
Matsubara, K.~(2013). `Realism, Underdetermination and String Theory Dualities. {\it Synthese,} 190, 471-489.\\
\\
Munkres, J.~R.~(2000). {\it Topology}. Prentice-Hall, Second Edition.\\
\\
Quine, W.~V.~(1970). `On the Reasons for Indeterminacy of Translation'. {\it The Journal of Philosophy,} 67 (6), pp.~178-183.\\
\\
Quine, W.~V.~(1975). `On empirically equivalent systems of the world'. {\it Erkenntnis}, 9 (3), pp.~313-328.\\
\\
Quine, W.~V.~(2002). `Epistemology Naturalized'. A reprint of Quine's 1971 paper. In: Brad Wray, K.~(Ed.), {\it Knowledge and Inquiry. Readings in Epistemology}.\\
\\
Read, J.~(2016). `The Interpretation of String-Theoretic Dualities'. {\it Foundations of Physics,} 46 (2), pp.~209-235.\\
\\
Read, J.~and M\o ller-Nielsen, T.~(2018). `Motivating Dualities'. Forthcoming in {\it Synthese}.\\
\\
Rickles, D.~(2017). `Dual theories: `Same but different' or `different but same'?' {\it Studies in History and Philosophy of Modern Physics,} 59, pp.~62-67.\\
\\
Savit, R.~(1980). `Duality in field theory and statistical systems'. {\it Reviews of Modern Physics,} 52 (2, I), pp.~453-487.\\
\\
van Fraassen, B.~C.~(1980). {\it The Scientific Image}. Oxford: Clarendon Press.\\
\\
van Fraassen, B.~C.~(1989). {\it Laws and Symmetry}. Oxford: Clarendon Press.\\
\\
van Fraassen, B.~C.~(2014). `One or Two Gentle Remarks about Hans Halvorson's Critique of the Semantic View'. {\it Philosophy of Science}, 81, pp.~276-283.\\
\\
Weatherall, J.~O.~(2016). `Are Newtonian gravitation and geometrized Newtonian gravitation theoretically equivalent?' {\it Erkenntnis}, 81 (5), pp.~1073-1091.\\
\\
Weatherall, J.~O.~(2019). `Equivalence and Duality in Electromagnetism'. Preprint, arXiv:1906.09699 [physics.hist-ph]. PhilSci 16149. https://arxiv.org/abs/1906.09699\\
\\
Weyl, H.~(1934). {\it Mind and Nature.} Reprinted in: Pesic, P.~(Ed.), {\it Hermann Weyl, Mind and Nature,} Selected Writings on Philosophy, Mathematics, and Physics, 2009. Princeton and Oxford: Princeton University Press.\\

\end{document}